\documentclass[%
 reprint,
 amsmath,amssymb,
 aps,
 prl,
floatfix,
]{revtex4-1}

\usepackage{graphicx}% Include figure files
\usepackage{dcolumn}% Align table columns on decimal point
\usepackage{bm}% bold math
\usepackage{physics}
\usepackage{color}
\begin{document}

%\preprint{APS/123-QED}

\title{Effective Modeling of Open Quantum Systems by Low-rank Discretization of Structured Environments}
%in a Linear Dissipative Environment}

% Force line breaks with \\
%\thanks{RB wisshes to thank...}%

\author{Hideaki Takahashi}
 \email{hideaki.takahashi@unito.it}
\author{Raffaele Borrelli}%
 \email{raffaele.borrelli@unito.it}
\affiliation{University of Torino, DISAFA, I-10095, Grugliasco, Italy}%

\date{\today}% It is always \today, today,
\begin{abstract}
The accurate description of the interaction of a quantum system with a its environment is a challenging problem ubiquitous across all areas of physics, and lies at the foundation of quantum mechanics theory. 
Here we pioneer a new strategy to create discrete low-rank models of the 
system-environment interaction, by exploiting the frequency and time domain information encoded in the fluctuation-dissipation relation
connecting the system-bath correlation function and the spectral density.
We demonstrate the effectiveness of our methodology by 
combining it with tensor-network methodologies and simulating the quantum dynamics of a complex excitonic systems in a highly structured bosonic environment. 
The new modeling framework sets the basis for a leap in the analysis of open quantum systems providing controlled accuracy at significantly reduced computational costs, with benefits in all connected research areas.
\end{abstract}

%Use showkeys class option if keyword
\maketitle

%\tableofcontents

%\section{\label{sec:level1}First-level}

%%%%%%%%%%%%%%%%%%%%%%%%%%%%%%%%%%%%%%%%%%%%%%%%%%%%%%%%
%%%%%%%%%%%%%%%% MOTIVATIONS
%%%%%%%%%%%%%%%%%%%%%%%%%%%%%%%%%%%%%%%%%%%%%%%%%%%%%%%
An incredibly rich variety of physical behaviours emerge from the interaction of a quantum system with its environment,\cite{Weiss2012QDS} and
their investigation and manipulation is one of the cornerstones of modern physical science, being intertwined with countless research  domains.
%Quantum electrodynamics,\cite{Carmichael1993} 
%
System-environment interactions play a prominent role 
in quantum phase transitions;\cite{BullaEtAl2003PRL,WinterEtAl2009PRL}
exciton and charge transport dynamics in organic electronics are fundamentally controlled by the vibrational environment of large molecular assemblies\cite{PoppEtAl2021ARPC}.
In quantum computing, environmental effects have destructive impact on the efficiency of quantum algorithms, mostly due to decoherence phenomena.\cite{PalmaEtAl1996PMPES}  
However, if properly rationalzied, this interaction can, in principle, be exploited to engineer strongly correlated states that can be used in quantum computations.\cite{VerstraeteEtAl2009NP,BeigeEtAl2000PRL}
Also, specific physical and chemical processes can be induced by the interaction of a system with environmental modes of a cavity quantum electromagnetic field.\cite{AberraGuebrouEtAl2012PRL,DuYuen-Zhou2022PRL}

In most cases of practical interest the environment comprises a continuous boson or fermion field, and its interaction with the system is characterized by the spectral density function  $J(\omega)$. 
%%%%%%%%%%%%%%%%%%%%%%%%%%%%%%%%%%%%%%%%%%%%%%%%%%%%%%%%
% Introduce BCF and SD and their use in quantum dynamics
%%%%%%%%%%%%%%%%%%%%%%%%%%%%%%%%%%%%%%%%%%%%%%%%%%%%%%%%
The cornerstone of many open quantum system theories is that the overall dynamical
effect of the environment is encoded in the so-called system-bath correlation
function (BCF), $C(t)$\cite{BreuerPetruccione2010,
CallenWelton1951PR,Kubo1957JPSJ,FeynmanVernon1963AP}
which is related to $J(\omega)$ through a particular form of fluctuation-dissipation relation (FDR)\cite{Kubo1966RPP} which, for a boson environment, is 
\begin{equation}
    C(t) =\frac{1}{2}\int_{-\infty}^{+\infty}\!\!\!\!\dd\omega
    J(\omega)\left[\coth(\beta\omega/2)+1\right]e^{-i\omega t},\label{eq:bcf}
\end{equation}
where 
$\beta$ is the inverse temperature of the environment, and 
$J(\omega)$ is assumed to be an odd function.\cite{KuboStatPhysII1991} 
In the following we will focus on bosonic environment, but the theoretical and computational framework applies to fermionic fields as well.

%%%%%%%%%%%%%%%%%%%%%%%%%%%%%%%%%%%%%%%%%%%%%%%%%%%%%%%%%%
%%%%%%%%%%% WHATS IS THE PROBLEM PART 1
%%%%%%%%%%%%%%%%%%%%%%%%%%%%%%%%%%%%%%%%%%%%%%%%%%%%%%%%%%
When system-environment correlations are strong or the characteristic memory time 
of the environment is of the same order of the time evolution of the system,
analytical approaches do not provide qualitatively correct results, and numerical modeling tools are pivotal for studying open quantum systems dynamics. 
It is therefore not surprising that an impressive number of numerical methodologies have been developed for this purpose.\cite{
MeyerEtAl2009,WangThoss2003JCP,Tanimura2020JCP,
Zhao2023JCP,Makri1995JMP,BorrelliGelin2016JCP,ChinEtAl2013NP,LorenzoniEtAl2024PRL,CygorekEtAl2022NP,StrathearnEtAl2018NC}
In order to capture the intricate 
system-environment dynamics, most approaches relies on the discretization of the continuous boson field, on a set of frequencies $\{\omega_k\}$.
The pairs of values $\{\omega_k,J(\omega_k)\}$ are then used to build a microscopic Hamiltonian operator by mapping the environment onto a set of harmonic oscillators 
coupled to the system, resulting in the so-called system-bath Hamiltonian.\cite{LeggettEtAl1987RMP}
Among others, sophisticated models of exciton and charge-transport processes have been built in the last three decades using this approach,\cite{MakriEtAl1996PNAS, BurghardtEtAl2008JCP,SchnedermannEtAl2019NC} providing a significant contribution to the understanding of their fundamental mechanisms. 
Moreover, the introduction of Tensor-Networks methodologies to describe complex and entangled vibronic wavefunctions has boosted and extended the adoption and development of this type of models to unprecedented levels.\cite{SchroderEtAl2019NC,PoppEtAl2021ARPC,ChinEtAl2013NP,KeEtAl2022JCP,WangEtAl2011JCP,WangThoss2013JCP,KunduMakri2020JPCL}

When the environment is discretized into a finite number of modes, $M$, the BCF takes the form 
\begin{equation}
    C(t) = \sum_{k=1}^M g_k^2(\beta) e^{-i\omega_k t}. \label{eq:bcf-disc}
\end{equation}
where the $g_k(\beta)$ are temperature dependent real parameters. 
%Furthermore, this approach has recently been recently extended to address finite temperature effects by the methods of thermo-field dynamics.
Using the theoretical framework of thermo-field dynamics it is possible to demonstrate that the system-environment dynamics driven by the BCF Eq. (\ref{eq:bcf-disc}), is controlled by the generalized Hamiltonian
operator
%In this framework, it can be demonstrated that the dynamics driven by the BCF Eq. (\ref{eq:bcf-disc}) can be directly mapped onto the generalized Hamiltonian operator  
\cite{BorrelliGelin2016JCP,GelinBorrelli2017AdP,BorrelliGelin2017SR,GelinEtAl2021JCP,BreyEtAl2021JPCC,TamascelliEtAl2019PRL,deVegaBanuls2015PRA} 
\begin{equation}
    H = H_\mathrm{S} + \sum_{k=1}^M\omega_k a_k^\dagger a_k + V_\mathrm{SB}\sum_{k=1}^M g_k(\beta) (a_k^\dagger + a_k)\label{eq:sbmodel}
\end{equation}
where $H_\mathrm{S}$ is the system Hamiltonian, $a^\dagger_k,a_k$ are creation-annihilation operators of an extended bosonic environment, and $V_{SB}$ is a system operator describing the interaction with the environment. 
A key feature of the above finite-temperature model is that the frequencies $\omega_k$ can assume both positive and negative values, and the system-environemnt coupling depends on the temprature.
%\:In thermo-field dynamics formalism modes associated with negative frequencies belong to the so-called \textit{tilde} space. 
%More details on the TFD theory are given in the Supplemental Material (SM).
See the Supplemental Material (SM) for more details on finite-temperature Hamiltonian.
%%

%%%%%%%%%%%%%%%%%%%%%%%%%%%%%%%%%%%%%%%%%%%%%%%%%%%%%%%%%%%%
%%%%%%%%% Whats is the problem part 2: the real point
%%%%%%%%%%%%%%%%%%%%%%%%%%%%%%%%%%%%%%%%%%%%%%%%%%%%%%%%%%%%
The ability to simulate the overall system-environment dynamics depends, first and foremost, on the capability 
to develop numerically convergent models, in which
the set of parameters $\{\omega_k, g_k(\beta)\}$, 
provides an accurate description of the BCF, yet
its size $M$ is small enough to  
be handled by current quantum dynamical methodologies.
%%
%%%%%%%%%%%%%%%%%%%%%%%%%%%%%%%%%%%%%%%%%%%%%%%%%%%%%%%%%%%%
%%%%%%%%%%%%%%%%%%  Available procedures
%%%%%%%%%%%%%%%%%%%%%%%%%%%%%%%%%%%%%%%%%%%%%%%%%%%%%%%%%%%%
Simple equispaced sets of frequencies are very common\cite{TamuraBurghardt2013JACS,BurghardtEtAl2008JCP}
but certainly far from being optimal,\cite{BurkeyCantrell1984JOSABJ} while 
logarithmic distribution has been used for special cases of Ohmic or sub-Ohmic spectral densities\cite{BullaEtAl2003PRL}. 
More general discretization procedures 
based on a renormalization of the reorganization 
energy \cite{WangEtAl2001JCP,Makri1999JPCB}
and on the use of classical correlation functions\cite{WaltersEtAl2017JCC} have
have also been very frequently used.
Alternatively, it is possible to map the linear model of Eq. (\ref{eq:sbmodel}) into a chain of nearest-neighbour interacting particles
\cite{CederbaumEtAl2005PRL,
ShenviEtAl2008PRA,deVegaEtAl2015PRB,
PriorEtAl2010PRL,ChinEtAl2010JoMP,TamascelliEtAl2019PRL} with the advantage that, in principle, a relatively small number of degrees of freedom (DoF) should be sufficient to reproduce the short time dynamics.\cite{CederbaumEtAl2005PRL} 

%%%%%%%%%%%%%%%%%%%%%%%%%%%%%%%%%%%%
%%%% THE REAL CHALLENGE  %%%%%%%%%
%%%%%%%%%%%%%%%%%%%%%%%%%%%%%%%%%%%%
However, developing a methodology that can provide both optimal sampling frequencies $\{\omega_k\}$ and coupling coefficients $\{g_k(\beta)\}$, valid for any arbitrary structured spectral density at a specified \textit{a priori} accuracy, remains a challenge.
%is still missing and would be extremely valuable.

Here we provide such a method, and demonstrate its 
practical use in the simulation of the exciton dynamics in a highly structured bosonic environment.
We pioneer a new paradigm in the construction of discrete models, whose key idea is the use of
%At the core of the method is the use of
a special low--rank approximation to compress the time, frequency and temperature information encoded in the FDR Eq. (\ref{eq:bcf}), and reduce it to the form of Eq. (\ref{eq:bcf-disc}), whereby the discrete model of Eq. (\ref{eq:sbmodel}) is recovered.
Specifically, we  cast the FDR into a matrix equation and employ the theory of Interpolative Decomposition (ID) of matrices to efficiently select the relevant environment frequencies and generate optimized coupling parameters, $g_k(\beta)$. \cite{ChengEtAl2005SJSC,LibertyEtAl2007PNAS,MartinssonEtAl2007CAMCS} 

We begin by writing the FDR in the discretized form
\begin{equation}
    C(t_i) = \sum_{j=1}^n f(t_i,\omega_j) w_j +e_i \label{eq:BCFID}
\end{equation}
where the $f(t_i,\omega_j)$ is the integrand of Eq. (\ref{eq:bcf}) evaluated at a specific time $t_i$ and frequency $\omega_j$, $w_j$ are some not yet specified weights,
and $e_i$ is an error term.
The error can be made arbitrarily small by properly choosing the grid points and the weights. 
Here we define $(t_i,\omega_j)$ to be the lattice coordinates of a dense $m\times n$ rectangular grid in the interval $[0,T]\times[-\Omega,\Omega]$, 
where $T$ is an upper time limit and $\Omega$ a given frequency value.
These two parameters must be properly chosen to capture all the relevant dynamics of interest. 
Usually $T$ is the upper time limit of interest in the dynamical problem, and $f(t,\omega) = 0$ for $\omega$ outside the frequency interval $[-\Omega,\Omega]$.
As we shall briefly see the time information is crucial for a proper application of the procedure because it is strongly affects the selection of the frequency set.
 
The set of points $f(t_i,\omega_j)$ is then reshaped into a $m\times n$ matrix  $f=(f_{ij} := f(t_i,\omega_j))$. 
The key point is realizing that the the matrix equation \ref{eq:bcf-disc} gives the vector $C$ as a linear superposition of a very large set of vectors $f$.
We thus seek for the best basis vectors $f$ to
represent the vector $C$.
To this end we use ID 
to determine a compressed (lossy) version of the matrix $f$ 
in the form of a product of two low--rank matrices\cite{LibertyEtAl2007PNAS}
\begin{equation}
     f_{m\times n} \approx B_{m\times r} P_{r\times n} \label{eq:LOWRANKID}.
\end{equation}
Here, $B$ is an $m \times r$ matrix consisting of $r$ columns selected from $f$, with $r \leq n$.
These columns are chosen so that they provide a good basis for approximating the entire matrix $f$. $P$ is a $r \times n$ matrix,
whose columns contain a mixture of 1s and 0s that 
effectively select the corresponding columns in $B$, and 
the other entries are coefficients that interpolate the 
remaining columns of $f$. Numerical methodologies for 
obtaining the decomposition of Eq. (\ref{eq:LOWRANKID}) to a 
desired accuracy, are described, for example, in references  \cite{LibertyEtAl2007PNAS,KayeEtAl2022PRB,KayeEtAl2022CPC}. See also the SM.

Once the ID of $f$ is performed, a set of columns of the original matrix are obtained that represents the function $f$ evaluated on a subset of frequencies $\{\omega_k\}$.
Thus, we can rewite Eq. (\ref{eq:BCFID}) in vector form
\begin{equation}
    %C \approx B P w + E = B z + E
    C \approx B P w = B z 
    \label{eq:BCFIDVEC}
\end{equation}
where $C=(C(t_i))$, $z = Pw$ is a new set of $r$ weights.
%Since the vector $C$ and the matrix $B$ are known, 
The coefficients $z_k$ can then be obtained by solving Eq. (\ref{eq:BCFIDVEC}) with the constraint $z_k\ge 0$ (see below). 
Since the columns of $B$, are taken directly from the columns of the original matrix $f$ we can rewrite Eq. (\ref{eq:BCFIDVEC}) in the explicit form
\begin{equation}
    C(t) \approx \sum_{k=1}^M z_k S_\beta(\omega_k)e^{-i\omega_k t} ~~ \mathrm{for}~~ t\in [0,T], \label{eq:BCFapprox} 
\end{equation}
where $S_\beta(\omega_k)=\frac{1}{2}J(\omega_k)(\coth(\beta\omega_k/2)+1)$, 
and the approximation holds in the time interval $[0,T]$.

%We notice that the determination of $z$ is an overdetermined problem, therefore 
%approximate solutions via non-negative least square\cite{LawsonHanson1995SLSP} 
%(NNLS) procedure or convex optimization methods are required.\cite{BoydVandenberghe2004} In the following
%we use the NNLS approach for its robustness.
%We further note that, from a numerical point of view, the matrix $P$ is absorbed 
%into the coefficients $z_k$ and never explicitly used.
We point out that ID induces a pivoting of the columns of $f$, thus the $M$ frequencies $\{\omega_k\}$ do not correspond to the first $M$ frequencies of the initial  discretization grid, but are a special subset of the latter.  
Eq. (\ref{eq:BCFapprox})  is directly mapped onto the generalized Hamiltonian given in Eq. (\ref{eq:sbmodel}) with
\begin{equation}
    g_k(\beta) = \sqrt{z_k S_\beta(\omega_k)}, \label{eq:GJBETA}
\end{equation}
where we notice that, for the Hamiltonian operator to be Hermitian,  the weights $z_k$ must be non-negative. To this end we solve Eq. (\ref{eq:BCFIDVEC}) using Non-Negative Least Squares (NNLS).\cite{LawsonHanson1995SLSP}
A key aspect of our approach is that the coupling parameters $g_k(\beta)$ depend not only on the value of $S_\beta(\omega_k)$, as in all discretization procedures proposed so far in the literature, but also on the optimized weights $z_k$. This additional flexibility is provided by the time domain information encoded in $C(t)$, and, together with ID, it is at the root of the efficiency of the method.
The number of terms $M$ in the expansion Eq. (\ref{eq:BCFapprox}), is determined by \textit{i}) the accuracy  of the ID and \textit{ii}) the accuracy of the NNLS solution for the weights $z_k$.
Finally, we note that, because the NNLS procedure can result in some of the coupling coefficients $z_k$ to be zero, $M \leq r$. 

%%%%%%%%%%%%%%%%%%%%%%%%%%%%%%%%%%%%%%%%%%%%%%%%
%%%%%%%%%%%%%%%%%% RESULTS
%%%%%%%%%%%%%%%%%%%%%%%%%%%%%%%%%%%%%%%%%%%%%%%%
In order to show the actual potential of our approach we use it to construct a vibronic model for the exciton transfer in the Fenna-Mathews-Olsen (FMO) complex at 0 K, 77 K and 300 K, based on the
experimentally determined spectral density reported in ref. \cite{WendlingEtAl2000JPCB}.
This model has quickly become a test bed for many open quantum dynamics theories.\cite{MoixEtAl2011JPCL,MoixEtAl2012PRB,
SchulzeKuhn2015JPCB,SchulzeEtAl2016JCP,BorrelliGelin2017SR}
%is first refined using a smoothed interpolating spline to remove part of the background noise. 

\begin{figure}
    \centering
    \includegraphics[width=8.2cm]{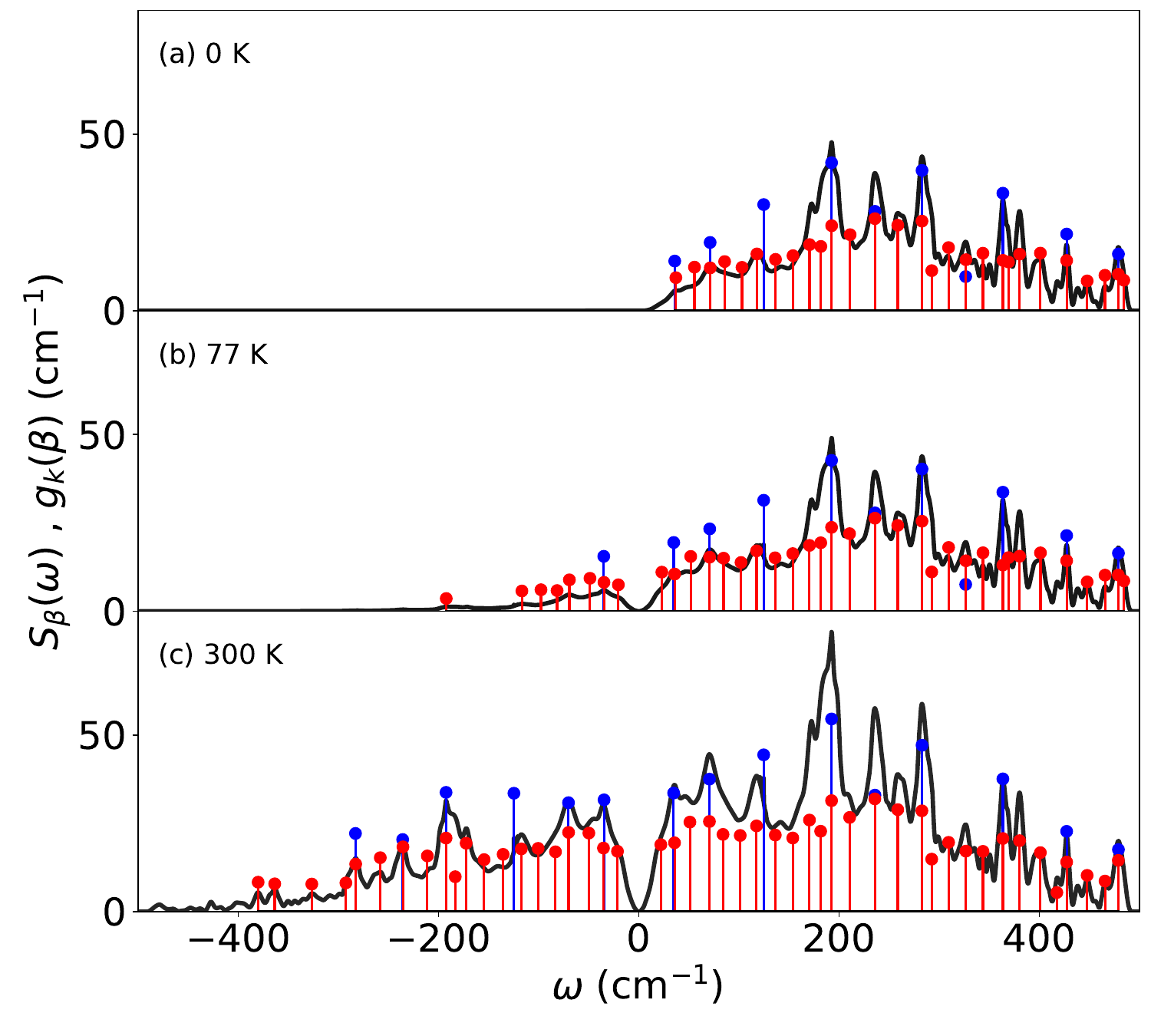}
    \caption{Effective spectral density $S_\beta(\omega)$ for the FMO model system (black), coefficients $g_k(\beta)$ and their corresponding frequencies $\omega_k$ 
    obtained using $T=1000$ fs (red bars), and $T=300$ fs (blue bars).
    %Blue bars:
    (a) 0 K, $M=10$ (blue), $M=28$ (red) (b) 77 K, $M=11$ (blue), $M=37$ (red), and (c) 300 K, $M=15$ (blue), $M=48$ (red) for $T=300$ fs. 
    %Red bars:  $T=1000$ fs, (a) 0 K, $M=28$, (b) 77 K, $M=37$ and (c) 300 K, $M=48$.  
    }
    \label{fig:fmosbeta}
\end{figure}

In figure \ref{fig:fmosbeta} we plot the quantum noise function $S_\beta(\omega)$ 
at different temperatures, and the effective coupling coefficients $g_k(\beta)$
at the corresponding frequency $\omega_k$, as obtained from the ID methodology.
The red and blue vertical lines show the result obtained employing two different rectangular grids with $T=1000$ fs and $T=300$ fs, respectively. In both cases, the grids comprises 1000 points in time and 10000 points in frequency domain. The discretized sets correspond to an average error in the BCF of about $10^{-2}$ (see SM).
We immediately observe that  the number of sampling points is significantly smaller for $T=300$ fs compared to $T=1000$ fs. 
Stated differently, far fewer DoFs are needed to describe the dynamics for $T=300$ fs than for $T=1000$ fs.
This is a much desired behaviour, and a distinctive feature of the methodology that is capable of exploiting both the frequency and the time domain information of the FDR.
%%%%
%%%
\begin{figure}
    \centering
    \includegraphics[width=8.2cm]{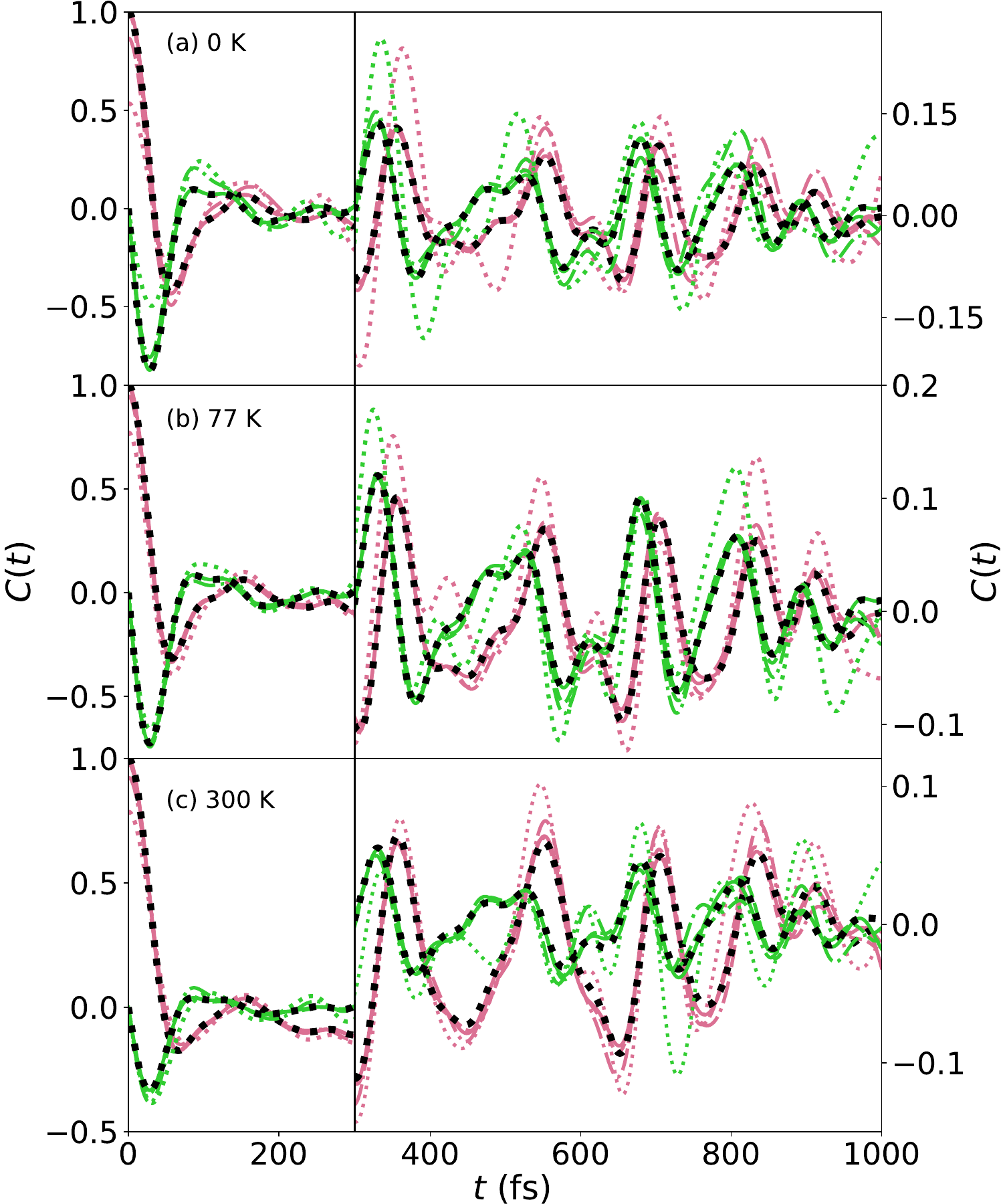}
    \caption{Real part (pink) and imaginary part (green) of FMO System-bath correlation function at (a) 0 K, (b) 77 K and (c) 300 K approximated with ID using different numbers of sample points: 
    (a) $M=9$ (dotted), 
    $19,$ (dashdot), 
    $28$ (dashed) and 
    $33$ (solid), 
    (b) $M=19$ (dotted), 
    $29$ (dashdot), 
    $37$ (dashed) and 
    $48$ (solid) and (c) 
    $M=29$ (dotted), $39$ (dashdot), 
    $48$ (dashed) and 
    $56$ (solid).  
    Black dotted lines are the references.
     All calculations have been performed setting $T=1000$ fs.}
    \label{fig:fmobcf}
\end{figure}
%%%%
\begin{figure}
    \centering
    \includegraphics[width=8.2cm]{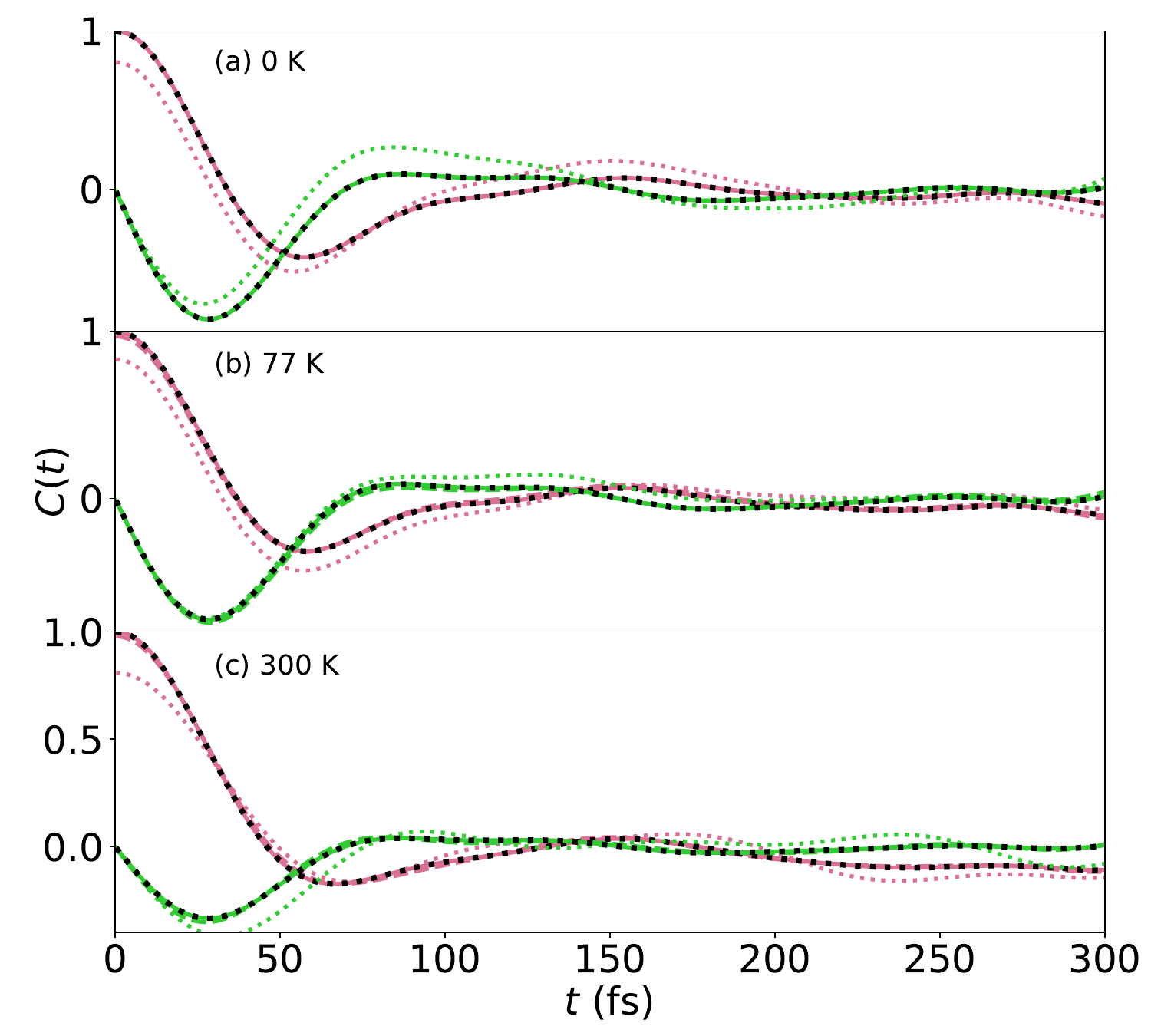}
    \caption{Real part (pink) and imaginary part (green) of FMO System-bath correlation function at (a) 0 K, (b) 77 K and (c) 300 K approximated with ID using different numbers of sample points: (a) $M=6$ (dotted), $10$ (dashed) and $12$ (solid), (b) $M=7$ (dotted), $11$ (dashed) and $15$ (solid) and (c) $M=10$ (dotted), $15$ (dashed) and $19$ (solid).  Black dotted lines are the references. All calculations have been performed setting $T=300$ fs.}
    \label{fig:fmobcf300}
\end{figure}
%% Comment figure
%From the point of view of the BCFs analysis of the procedure can be obtained from a comparison of

The compressed BCFs are shown in
figures \ref{fig:fmobcf}, and \ref{fig:fmobcf300}, 
for $T=1000$ fs and $T=300$ fs, respectively, with increasing value of $M$.
In figure \ref{fig:fmobcf} the scale of the vertical axis is changed at $t=300$ fs, for clarity. 
At 0 K, accurate fitting of the BCF up to 1000 fs is already obtained with $M=28$, which become 37 at 77 K, and 48 at 300 K. 
The reference value has been computed directly using Eq. (\ref{eq:BCFID}) on a very dense grid.
Figure \ref{fig:fmobcf300} shows the same behaviour, \textit{i.e.}  in all cases the accuracy increases with the number of expansion terms. However, in this latter case the overall
number of expansion terms is much smaller due the decreased time interval. At 300 K, the converged BCF can be obtained by using only 15 terms, for example. 

\begin{figure}
    \centering
    \includegraphics[width=8.2cm]{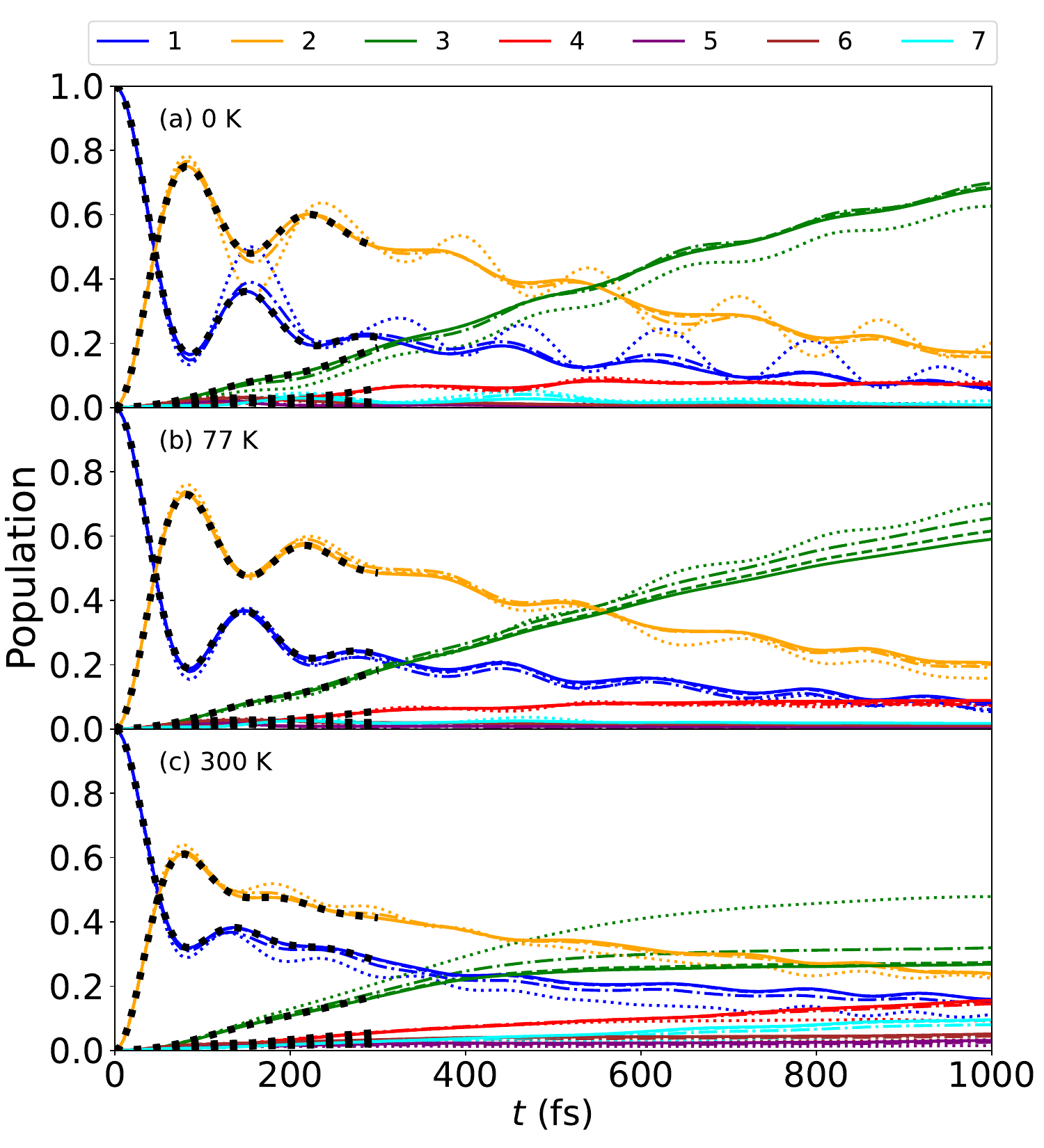}
    \caption{Site populations at (a) 0 K, (b) 77 K and (c) 300 K, of a seven-site model of the FMO using the Hamiltonian operator obtained from the experimental spectral density after ID using different numbers of sample points: (a) $M=9$ (dotted), 
    $19$ (dashdot), 
    $28$ (dashed) and 
    $33$ (solid), 
    (b) $M=19$ (dotted), 
    $29$ (dashdot), 
    $37$ (dashed) and 
    $48$ (solid) and (c)
    $M=22$ (dotted), 
    $39$ (dashdot), 
    $48$ (dashed) and 
    $56$ (solid).  Black dots represent the converged population dynamics using the BCF determined with $T=300$ fs.  All calculations are converged using Tensor-Train cores with rank $80$ for $T=300$ fs and $110$ for $T=1000$ fs.
    }
    \label{fig:fmo_pop}
\end{figure}

To test the validity of the new decomposition methodology, we
use it to construct a series of convergent Hamiltonian operators of the form of Eq. (\ref{eq:sbmodel}) 
for a seven-site FMO model at different temperatures. The resulting finite-temperature Schr\"odinger equation is then solved using the Tensor-Train/Matrix Product State representation of the wave function, and the 
time-dependent variational principle (TDVP) integrator.\cite{BorrelliGelin2016JCP,BorrelliGelin2021WCMS,
LubichEtAl2015SJNA,Oseledets2011SJSC} 
Figure \ref{fig:fmo_pop} shows the site populations as a function of time, for different temperatures, and for different number of DoFs. 
The colored lines represent the results based on the BCF decomposition obtained with $T=1000$ fs.
%%%
%%%%
First of all, our results reveal that with increasing
the number of discretization points $M$, 
the population curves converge uniformly across the entire time domain. 
The black dots represent the converged dynamics obtained for $T=300$ fs.
In this case the model Hamiltonian contains only 15 DoFs per molecule.
We observe that the two sets of results overlap perfectly.
This confirms that the information contained in the 
FDR is fundamental for effectively sampling the frequency domain, and
underscores the reliability and robustness of the methodology.

Notably, the dynamics of the model at 300 K is converged up to 1000 fs using only 48 DoFs per molecule. 
%This efficiency remains effective even at lower temperatures, 
%allowing for economical yet accurate modelling across a wide range of thermal conditions.
This is of particular relevance for numerical simulations of open systems. 
Indeed,  we note that previous works employed a uniform set of frequencies
with a sampling interval of about 4 $\mathrm{cm}^{-1}$,
\cite{SchulzeKuhn2015JPCB,SchulzeEtAl2016JCP,BorrelliGelin2017SR} 
which results in a model with 250 DoFs per site, if applied to the present $S_\beta(\omega)$ in the range $[-500,500]$ cm$^{-1}$. 
The capability of the methodology to grasp the essential dynamical features of the environment is thus quite evident.
With a more than five-fold decrease in the number of DoFs of the environment, the computational saving introduced by our approach is a real breakthrough.

%%%%%%%%%%%%%%%%%%%%%%%%%%%%%%%%%%%%%%%%%%%%%%%%%%%%%
%%%%%%% Conclusions; what we addressed 
%%%%%%%%%%%%%%%%%%%%%%%%%%%%%%%%%%%%%%%%%%%%%%%%%%%%%
In summary, we have proposed a new comprehensive methodology for developing numerically convergent microscopic models of a quantum system interacting with an arbitrary structured environment valid at any temperature. 
The approach relies on the application of the ID theory  for the determination of a low-rank approximation of 
the fluctuation-dissipation relation connecting the BCF and the spectral density function,
and can exploit both frequency and time information encoded therein.
We have shown that by combining this novel approach with modern tensor-network-based tools for quantum dynamics simulations, 
it is possible to  effectively describe the exciton transfer dynamics in a large molecular complex.
Our results clearly demonstrate the potential of the method, its accuracy, and the significant computational savings achieved compared to existing methodologies. 
We have presented an application to a bosonic dissipative environment, however, we remark that this approach is absolutely general, and can be applied to fermionic reservoirs as well.
%We also envisage the extension of this methodology to non-Gaussian bath, and reservoirs of anharmonic oscillators.\cite{HsiehCao2018TJoCP,HsiehCao2018TJoCPa} 
%This could be achieved by applying ID to higher order correlation functions which provide information on the most relevant anharmonicities.
We believe that this new methodology can be incredibly valuable, especially for all the applications where precise spectral information is crucial, and
opens up new possibilities for simulating complex quantum dynamics in a variety of research areas.

\begin{acknowledgments}
The authors acknowledge the support by the Spoke 7 "Materials and Molecular Sciences" of ICSC – Centro Nazionale di Ricerca in High-Performance Computing, Big Data and Quantum Computing, funded by European Union – NextGenerationEU. 
R.R. acknowledge the support of the project “nuovi Concetti, mAteriali e tecnologie per l’iNtegrazione del fotoVoltAico negli edifici in uno scenario di generazione diffuSa” [CANVAS], funded by the Italian Ministry of the Environment and the Energy Security, through the Research Fund for the Italian Electrical System (type-A call, published on G.U.R.I. n. 192 on 18-08-2022)
\end{acknowledgments}

\bibliography{mylibrary,book}
% Produces the bibliography via BibTeX.

\end{document}